%% file: letters.tex
\documentclass[journal]{IEEEtran}
\usepackage{cite}
\usepackage{amssymb}
\usepackage[pdftex]{graphicx}
\usepackage{amsfonts}
\usepackage{lineno}

\input latex_defs.tex

\usepackage[cmex10]{amsmath}
\usepackage{array}
\usepackage{stfloats}
\usepackage{url}
\usepackage{epstopdf}
\usepackage[square,numbers,sort&compress]{natbib}
\usepackage{subfigure}
\usepackage{subeqnarray}
\usepackage{cases}

\input latex_defs.tex
\allowdisplaybreaks
\def\j{\textrm{j}}

\input latex_defs.tex
\allowdisplaybreaks
\begin{document}
\title{TE and TM Modes in Metallic Waveguide Filled With a Lossless and Fully Anisotropic Medium}
\author{\IEEEauthorblockN{Wei Jiang and Jie Liu}

\thanks{This work was supported by the National Natural Science Foundation of China under Grant 61901131, the Natural Science Foundation of Guizhou Minzu University under Grant GZMU[2019]YB07 and the China Postdoctoral Science Foundation under Grant 2019M662244 \emph{(Corresponding author: Wei Jiang).}}
\thanks{W. Jiang is with the School of Mechatronics Engineering, Guizhou Minzu University, Guiyang 550025, China (e-mail: jwmathphy@163.com).
}
\thanks{J. Liu is with the Postdoctoral Mobile Station of Information and Communication
Engineering, School of Informatics, and Institute of Electromagnetics
and Acoustics, Xiamen University, Xiamen 361005, China (e-mail: liujie190484@163.com).
}}
%\markboth{IEEE MICROWAVE AND WIRELESS COMPONENTS LETTERS}%
\markboth{}%
{Shell \MakeLowercase{\textit{et al.}}: Wei Jiang.}

\maketitle

\begin{abstract}
This letter gives a condition for the existence of the pure TE and TM modes in the metallic waveguide filled with a homogeneous, fully anisotropic and lossless medium. The condition is a relation between the  permittivity and permeability tensors of the fully anisotropic medium. The theory given in the letter is the extension of the classic electromagnetic waveguide theory. At last, we employ the commercial software COMSOL Multiphysics to simulate the metallic waveguide problem filled with a homogeneous, fully anisotropic and lossless medium. Numerical experiment shows that the condition supported in the letter does confirm the existence of the pure TE and TM modes in the anisotropic waveguide.
\end{abstract}


%\IEEEpeerreviewmaketitle

\section{Introduction}
From the classic  waveguide theory in electromagnetics \cite{Balanis}, it is well known that there exist the pure TE and TM modes in the metallic waveguide filled with a homogeneous, isotropic and lossless medium.
Moreover, in such a metallic waveguide, the propagation mode is the pure TE and TM modes, instead of EH mode. If the metallic waveguide filled with a homogenous, isotropic and lossless medium has two independent perfect electric conductors at least, then TEM mode will exist in the waveguide. The pure TE mode can be simulated by the transversal components of electric field or the longitudinal component of magnetic field. Similarly, the pure TM mode can be simulated by the transversal components of magnetic field or the longitudinal component of electric field. For the TE mode, if the transversal components of electric field are applied to simulate it, then we have to solve an eigenvalue problem of the curl-curl operator with a Dirichlet boundary condition, which is the eigenvalue problem of a two-dimensional vector field. If the longitudinal component of magnetic field is used to simulate the TE mode, then we have to solve an eigenvalue problem of the Laplace operator with a Neumann boundary condition, which is the eigenvalue problem of a scalar field. For the TM mode, if we apply the transversal components of magnetic field to simulate it, then we have to solve an eigenvalue problem of the curl-curl operator with two Neumann boundary conditions, which is the eigenvalue problem of a two-dimensional vector field. If we employ the longitudinal component of electric field to simulate the TM mode, then we need to solve an eigenvalue problem of the Laplace operator with a Dirichlet boundary condition, which is the eigenvalue problem of a scalar field. The above discussion is the basic knowledge of classic electromagnetic waveguide.\\
\indent We now know that some media in nature exhibit anisotropic characteristics, for example, bamboo, magnetized ferrite material, uniaxial crystal, biaxial crystal and so on.
In recent years, Hu \emph{et.al} \cite{Hu2017} employ near-field optical technology to quantitatively determine the full dielectric tensors of nanometer-thin molybdenum disulfide and hexagonal boron nitride microcrystals, the most-promising van der Waals semiconductor and dielectric.
In addition, they theoretically demonstrate the existence of the pure TE and TM mode in the anisotropic van der Waals material. Therefore, it is worthwhile to propose the mode theory of the metallic waveguide filled with a homogeneous, anisotropic and lossless medium.
In general, the propagation mode in the anisotropic waveguide is EH mode, not the pure TE and TM modes because the coupling of electromagnetic fields in an anisotropic medium is very strong.
However, when the permittivity and permeability tensors of an anisotropic medium satisfy certain relationship, there are pure TE mode and TM mode in anisotropic waveguide.
For example, based on the Abelian group theory and the classic electromagnetic waveguide theory, Jiang \emph{et.al} \cite{Jiangq2018} establish a sufficient condition which can ensure the existence of the pure TE and TM modes in the anisotropic waveguide. On the basis of reference \cite{Jiangq2018}, Jiang \emph{et.al} \cite{Jiang2017n} propose a necessary and sufficient condition for the existence of the pure TE and TM modes in the anisotropic waveguide. The medium considered in \cite{Jiangq2018} and \cite{Jiang2017n} are only anisotropic in the plane of the waveguide cross section, and this is to say that the medium considered in \cite{Jiangq2018} and \cite{Jiang2017n} is not fully anisotropic. In this letter, to our surprising, the pure TE and TM modes may also exist in the metallic waveguide filled with a homogeneous, fully anisotropic and lossless medium when the permittivity and permeability tensors of the fully anisotropic medium satisfy certain relationship, which is given in this letter. \\
\indent There are several calculation methods to analyze the propagation mode in the waveguide.
For example, the variational meshless method in conjunction with radial basis functions can be used to solve the waveguide problem \cite{bardi2019}. Jiang \emph{et. al.} \cite{Jiang20188} make use of mixed finite element method to analyze the propagation mode in the metallic waveguide filled with anisotropic media. Furthermore, there are many literatures
about analyzing waveguide mode by edge element method, (please see \cite{Koshiba1992,Bardi1993,Zhou1996,Nuno1997} and references therein). In order to obtain the propagation mode in the metallic waveguide filled with fully anisotropic media, this letter simulates the waveguide problem by COMSOL Multiphysics, in which finite element method will be used.\\
\indent The outline of the letter is as follows. The governing equations for the waveguide problem is given in Section \uppercase\expandafter{\romannumeral2}. In Section \uppercase\expandafter{\romannumeral3}, we give a necessary condition for the existence of the pure TE and TM modes in the fully anisotropic waveguide. In Section \uppercase\expandafter{\romannumeral4}, a numerical experiment is carried out to verify our theoretical result.
\section{Governing Equations for Waveguide Problem}
\indent Let $\d{\ep}$ and $\d{\mu}$ be the permittivity and permeability tensors of an anisotropic medium, respectively. As usual, $\d{\ep}$ and $\d{\mu}$ can be written in the form of a matrix. Assume that the medium filled in the metallic waveguide is homogeneous, lossless and fully anisotropic, then $\d{\ep}$ and $\d{\mu}$ are both Hermitian \cite{Chew1990}. Therefore, $\d{\ep}$ and $\d{\mu}$ can be written by the following form
\begin{eqnarray}
  \d{\ep}=\left[
  \begin{array}{cc}
    \d{\ep}_{tt} & \bar{\ep}_{tz} \\
    \bar{\ep}_{tz}^{\dagger} & \ep_{zz} \\
  \end{array}
\right],\quad
\d{\mu}=\left[
  \begin{array}{cc}
    \d{\mu}_{tt} & \bar{\mu}_{tz} \\
    \bar{\mu}_{tz}^{\dagger} & \mu_{zz} \\
  \end{array}
\right] \label{mu}
\end{eqnarray}
where the superscript symbol $\dag$ represents the complex conjugate, $\d{\ep}_{tt}$ and $\d{\mu}_{tt}$ are $2\times2$ submatrices in the $\d{\ep}$ and $\d{\mu}$ respectively. Notice that $\d{\ep}_{tt}$ and $\d{\mu}_{tt}$ are also both Hermitian.

Suppose that the electromagnetic wave in the waveguide propagates along $+\^z$ direction. For a time-harmonic field with angular frequency $\omega>0$, we would like to seek several propagable physical modes in the waveguide:
\begin{equation}\label{guide1}
    \E(x,y,z)=\ee(x,y)\textrm{e}^{-\j k_{z} z},~\H(x,y,z)=\h(x,y)\textrm{e}^{-\j k_{z} z}
\end{equation}
In the expression (\ref{guide1}), $k_{z}$ is the propagation constant, which can be applied to characterize the propagation characteristics of electromagnetic fields in waveguide system. In the expression (\ref{guide1}), $\ee(x,y)$ and $\h(x,y)$ are the field distribution in the waveguide system, and they are independent with longitudinal coordinate $z$. Note that the vectors $\ee(x,y)$ and $\h(x,y)$ are composed of three components. They can be split into transverse components and longitudinal components, and that is
\begin{eqnarray}
\ee(x,y)=\ee_{t}(x,y)+\^ze_{z}(x,y)\label{guide2}\\
\h(x,y)=\h_{t}(x,y)+\^zh_{z}(x,y)\label{guidet2}
\end{eqnarray}
In order to analyze the propagable modes in the waveguide, three-dimensional del operator $\nabla$ needs to be split into transverse components and longitudinal component, and that is
\begin{eqnarray} \nabla=\nabla_{t}+\^z\pd{}{z},~~~\nabla_{t}=\^x\pd{}{x}+\^y\pd{}{y}.\label{guidddde2}
\end{eqnarray}

Since there is no source distribution in the waveguide, then the electromagnetic field in the waveguide must satisfy the source-free Maxwell's equations
\begin{subequations} \label{eq:1}
\begin{numcases}{}
  \curl\E=-\j\omega\d{\mu}\H\mbox{~in~}\Gamma\label{eq:1a}\\
  \curl\H= \j\omega\d{\ep}\E\mbox{~in~}\Gamma\label{eq:1b}\\
 % \div(\d{\ep}\E)=0\mbox{~in~}\Gamma\label{qeq:1b}\\
%  \div(\d{\mu}\H)=0\mbox{~in~}\Gamma\label{peq:1b}\\
  \^n\times\E={\bf{0}}\mbox{~on~}\partial\Gamma\label{eq:1c}\\
  \^n\cdot(\d{\mu}\H)=0\mbox{~on~}\partial\Gamma\label{eq:1d}
\end{numcases}
\end{subequations}
where $\Gamma$ is the cross section of a given waveguide and
$\partial\Gamma$ is the boundary of the cross section $\Gamma$.
%In fact, the equations (\ref{qeq:1b}) and (\ref{peq:1b}) can be deduced from the equations (\ref{eq:1a}) and (\ref{eq:1b}) since $\omega$ is a known positive constant.
%However, we still list the equations (\ref{qeq:1b}) and (\ref{peq:1b}) in the source-free Maxwell's equations (\ref{eq:1}) because (\ref{qeq:1b}) and (\ref{peq:1b}) are very important to numerical simulations of the waveguide problem.

By substituting (\ref{mu})-(\ref{guidddde2}) into (\ref{eq:1a}) and (\ref{eq:1b}), one can obtain
the  following partial differential equations (PDEs):
%\begin{eqnarray}
% \curl\ee-jk_{z}(\^z\times\ee)&=&-j\omega\d{\mu}\h,~~\mbox{in}~\Gamma,\label{equ1}\\
% \curl\h-jk_{z}(\^z\times\h)&=&j\omega\d{\ep}\ee,~~\mbox{in}~\Gamma,\label{equ2}\\
% \div(\d{\ep}\ee)&=&\gamma(\d{\ep}\ee)\cdot\^z,~~\mbox{in}~\Gamma,\label{equ3}\\
% \div(\d{\mu}\h)&=&\gamma(\d{\mu}\h)\cdot\^z,~~\mbox{in}~\Gamma,\label{equ4}\\
% \^n\times\ee&=&0,~~\mbox{on}~\partial\Gamma,\label{equ5}\\
% \^n\cdot(\d{\mu}\h)&=&0,~~\mbox{on}~\partial\Gamma,\label{equ6}
%\end{eqnarray}
\begin{gather}
  \curlt\ee_{t} = -\j\omega(\bar{\mu}_{tz}^{\dagger}\h_{t}+\mu_{zz}h_z)\^z\mbox{~in~}\Gamma\label{eqn31} \\
  -\^z\times\nabla_{t}e_z-jk_{z}\^z\times\ee_{t}= -\j\omega(\d{\mu}_{tt}\h_{t}+\bar{\mu}_{tz}h_z)\mbox{~in~}\Gamma \label{eqn32}\\
  \curlt\h_{t} = \j\omega(\bar{\ep}_{tz}^{\dagger}\ee_{t}+\ep_{zz}e_z)\^z\label{eqn33}\mbox{~in~}\Gamma \\
  -\^z\times\nabla_{t}h_z-jk_{z}\^z\times\h_{t}=  \j\omega(\d{\ep}_{tt}\ee_{t}+\bar{\ep}_{tz}e_z)\mbox{~in~}\Gamma  \label{eqn34}
\end{gather}

A counterclockwise rotation matrix $R=
\begin{bmatrix}
0 & -1 \\
1 & 0 \\
\end{bmatrix}$ is introduced for the purpose of mode analysis, and the matrix $R$ exactly coincides with the rotation transformation $\^z\times$ in the $xy$-plane. After changing the $\^z\times$ in (\ref{eqn32}) and (\ref{eqn34}) into the matrix $R$, we can get the matrix system
\begin{subequations} \label{eq:2}
\begin{numcases}{}
  \j k_{z}\ee_{t}+j\omega R\d{\mu}_{tt}\h_{t}=-\nabla_{t}e_z-\j\omega R\bar{\mu}_{tz}h_{z}~~~\label{eq:2a}\\
  -\j\omega R\d{\ep}_{tt}\ee_{t}+\j k_{z}\h_{t}=\j\omega R\bar{\ep}_{tz}e_{z}-\nabla_{t}h_z~~~\label{eq:2b}
\end{numcases}
Define
$${\bf{f}}_{1}=-\nabla_{t}e_z-\j\omega R\bar{\mu}_{tz}h_{z},~{\bf{f}}_{2}=\j\omega R\bar{\ep}_{tz}e_{z}-\nabla_{t}h_z.$$
\end{subequations}
From (\ref{eq:2}), it is easy to obtain the following equations:
\begin{gather}
  (-k_{z}^2\d{I}_{2}-\omega^2R\d{\mu}_{tt}R\d{\ep}_{tt})\ee_{t}=\j k_{z}{\bf{f}}_{1}-\j\omega R\d{\mu}_{tt}{\bf{f}}_{2}\label{etezhz}\\
  (-k_{z}^2\d{I}_{2}-\omega^2R\d{\ep}_{tt}R\d{\mu}_{tt})\h_{t}=\j\omega R\d{\ep}_{tt}{\bf{f}}_{1}+\j k_{z}{\bf{f}}_{2}\label{htezhz}
\end{gather}
where $\d{I}_{2}$ is second-order identity matrix.
%here, the expressions of $\F_{1}$ and $\F_{2}$ are as follows:
%\begin{gather*}
%\F_{1}=-jk_{z}\gradt e_{z}+j\omega R\d{\mu}_{tt}\gradt h_{z}\\
%\quad\quad\quad\quad+\omega k_{z} R\bar{\mu}_{tz}h_{z}+\omega^2R\d{\mu}_{tt}R\bar{\ep}_{tz}e_{z} \\
%    \F_{2}=-jk_{z}\gradt h_{z}-j\omega R\d{\ep}_{tt}\gradt e_{z}\\
%\quad\quad\quad\quad -\omega k_{z} R\bar{\ep}_{tz}e_{z}+\omega^2R\d{\ep}_{tt}R\bar{\mu}_{tz}h_{z}
%\end{gather*}
Set
\begin{gather*}
\F_{1}=\j k_{z}{\bf{f}}_{1}-\j\omega R\d{\mu}_{tt}{\bf{f}}_{2},
~\F_{2}=\j\omega R\d{\ep}_{tt}{\bf{f}}_{1}+\j k_{z}{\bf{f}}_{2}\\
    \N_{1}=-k_{z}^2\d{I}_{2}-\omega^2R\d{\mu}_{tt}R\d{\ep}_{tt},~
    \N_{2}=-k_{z}^2\d{I}_{2}-\omega^2R\d{\ep}_{tt}R\d{\mu}_{tt}
\end{gather*}
The conclusion that $\det(\N_{1})=\det(\N_{2})$ can be proved by the basis knowledge in linear algebra. When $\det(\N_{1})=\det(\N_{2})\neq0$, then from
(\ref{etezhz}) and (\ref{htezhz}), one can get
\begin{gather}
\ee_{t}=\N_{1}^{-1}\F_{1},~~
\h_{t}=\N_{2}^{-1}\F_{2}.\label{realt2}
\end{gather}
Substituting (\ref{realt2}) into (\ref{eqn31}) and (\ref{eqn33}) respectively, we have
\begin{eqnarray}
\curlt(\N_{1}^{-1}\F_{1})= -\j\omega(\bar{\mu}_{tz}^{\dagger}\N_{2}^{-1}\F_{2}+\mu_{zz}h_z)\^z\label{hzez1} \\
\curlt(\N_{2}^{-1}\F_{2}) = \j\omega(\bar{\ep}_{tz}^{\dagger}\N_{1}^{-1}\F_{1}+\ep_{zz}e_z)\^z\label{hzez2}
%\curlt[\N_{1}^{-1}(-\gamma\gradt e_{z}+j\omega A\d{\mu}_{tt}\gradt h_{z}-j\omega \gamma A\bar{\mu}_{tz}h_{z}+\omega^2A\d{\mu}_{tt}A\bar{\ep}_{tz}e_{z})]\nonumber\\
%=-j\omega[\bar{\mu}_{tz}^{\dagger}\N_{2}^{-1}(-\gamma\gradt h_{z}-j\omega A\d{\ep}_{tt}\gradt e_{z}+j\omega\gamma A\bar{\ep}_{tz}e_{z}+\omega^2A\d{\ep}_{tt}A\bar{\mu}_{tz}h_{z})+\mu_{zz}h_z]\^z\\
%\curlt[\N_{2}^{-1}(-\gamma\gradt h_{z}-j\omega A\d{\ep}_{tt}\gradt e_{z}+j\omega \gamma A\bar{\ep}_{tz}e_{z}+\omega^2A\d{\ep}_{tt}A\bar{\mu}_{tz}h_{z})]\nonumber\\
%=j\omega[\bar{\ep}_{tz}^{\dagger}\N_{1}^{-1}(-\gamma\gradt e_{z}+j\omega A\d{\mu}_{tt}\gradt h_{z}-j\omega \gamma A\bar{\mu}_{tz}h_{z}+\omega^2A\d{\mu}_{tt}A\bar{\ep}_{tz}e_{z})+\ep_{zz}e_z]\^z
\end{eqnarray}
It is clear that (\ref{hzez1}) and (\ref{hzez2}) are only the PDEs for the longitudinal components $(e_{z},h_{z})$ of the electromagnetic field.
\section{The Condition for the Existence of the Pure TE and TM Modes}
The condition for the existence of the pure TE and TM modes is that (\ref{hzez1}) and (\ref{hzez2}) are decoupled with respect to the unknown functions $e_{z}$ and $h_{z}$.

From \cite{Jiang2017n}, when $\bar{\ep}_{tz}=\bar{\mu}_{tz}=0$, the sufficient and necessary condition is that $\N_{1}$ and $\N_{2}$ are two scalar matrices. As a consequence, if (\ref{hzez1}) and (\ref{hzez2}) are decoupled, then $\N_{1}$ and $\N_{2}$ must be two scalar matrices. Otherwise, the equations (\ref{hzez1}) and (\ref{hzez2}) will be coupled if $\N_{1}$ and $\N_{2}$ are not two scalar matrices, which shows that the propagation mode in the waveguide is EH mode, instead of the pure TE and TM modes. When $\N_{1}$ and $\N_{2}$ are two scalar matrices, then $\N_{1}=\N_{2}$ is valid \cite{Jiang2017n}.
In this case, it is very easy to know $-\omega^2A\d{\mu}_{tt}A\d{\ep}_{tt}=-\omega^2A\d{\ep}_{tt}A\d{\mu}_{tt}$ are also two scalar matrices. Set
\begin{eqnarray}
 -\omega^2R\d{\mu}_{tt}R\d{\ep}_{tt}=-\omega^2R\d{\ep}_{tt}R\d{\mu}_{tt}=k^2\d{I}_{2}\label{rr}\\
    \N_{1}=\N_{2}=(k^2-k_{z}^2)\d{I}_{2}=k_{t}^2\d{I}_{2},\label{dd}
\end{eqnarray}
where $k$ and $k_{t}$ are called the medium and cut-off wavenumbers in an anisotropic and lossless medium, respectively. It is impossible to define the medium and cut-off wavenumbers in a general anisotropic medium, in which (\ref{rr}) and (\ref{dd}) are not valid.

Suppose that $\N_{1}$ and $\N_{2}$ are two scalar matrices, then $\N_{1}=\N_{2}$. In this case, PDEs (\ref{hzez1}) and (\ref{hzez2}) can be reduced to the following PDEs:
\begin{eqnarray}
\curlt\F_{1}= -\j\omega(\bar{\mu}_{tz}^{\dagger}\F_{2}+\mu_{zz} h_z k_{t}^2)\^z\label{chzez1} \\
\curlt\F_{2} = \j\omega(\bar{\ep}_{tz}^{\dagger}\F_{1}+\ep_{zz} e_z k_{t}^2)\^z\label{chzez2}
\end{eqnarray}

For the pure TM modes, we need to set $h_{z}=0$ in (\ref{chzez1}) and (\ref{chzez2}), and then obtain
\begin{align}
&\quad~\curlt(R\d{\mu}_{tt}R\bar{\ep}_{tz}e_{z})\nonumber\\
&=[\bar{\mu}_{tz}^{\dagger}(- R\d{\ep}_{tt}\gradt e_{z}+\j k_{z}R\bar{\ep}_{tz}e_{z})]\^z\label{rerty5}\\
&\quad\curlt(-R\d{\ep}_{tt}\gradt e_{z}+\j k_{z}R\bar{\ep}_{tz}e_{z})\nonumber\\
&=[\bar{\ep}_{tz}^{\dagger}(-\j k_{z}\gradt e_{z}+\omega^2R\d{\mu}_{tt}R\bar{\ep}_{tz}e_{z})+\ep_{zz}e_zk_{t}^2]\^z\label{rerty6}
\end{align}

The equation (\ref{rerty6}) together with boundary condition can be applied to simulate the pure TM mode.
Obviously, (\ref{rerty5}) must be a trivial equation. Hence, we have
\begin{eqnarray}
&&\bar{\mu}_{tz}^{\dagger}R\bar{\ep}_{tz}=0\label{eye1}\\
&&(\d{\mu}_{tt}R\bar{\ep}_{tz})^{T}=-\bar{\mu}_{tz}^{\dagger}R\d{\ep}_{tt}\label{eye2}
\end{eqnarray}
where the superscript $T$ stands for the transpose of a given matrix.
\noindent Similarly, if the pure TE modes exist in the waveguide, one can get
\begin{eqnarray}
\bar{\ep}_{tz}^{\dagger}R\bar{\mu}_{tz}&=&0\label{eye3}\\
(\d{\ep}_{tt}R\bar{\mu}_{tz})^{T}&=&-\bar{\ep}_{tz}^{\dagger}R\d{\mu}_{tt}\label{eye4}
\end{eqnarray}
When $\d{\ep}$ and $\d{\mu}$ are two real matrices, it is very easy to prove that (\ref{eye1})-(\ref{eye2}) are valid if and only if (\ref{eye3})-(\ref{eye4}) are valid.

We take $\d{\ep}$ and $\d{\mu}$ as the following way:

\begin{equation}\label{EEE}
    \d{\ep}=\begin{bmatrix}
            a_{11}&a_{12}&a_{13}\\
            a_{12}&a_{22}&a_{23}\\
           a_{13}&a_{23}&a_{33}
    \end{bmatrix}\nu_{1},~~
        \d{\mu}=\begin{bmatrix}
            a_{11}&a_{12}&a_{13}\\
            a_{12}&a_{22}&a_{23}\\
           a_{13}&a_{23}&b_{33}
    \end{bmatrix}\nu_{2}
\end{equation}
\begin{figure}[ht]
  \centering
  {\subfigure[]{
    \label{terecreal1}
   \includegraphics[width=0.48\columnwidth,draft=false]{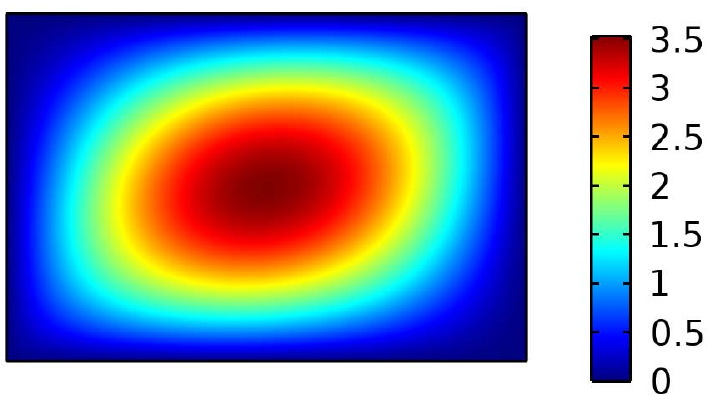}}
     \subfigure[]{
    \label{terecreal2}
   \includegraphics[width=0.48\columnwidth,draft=false]{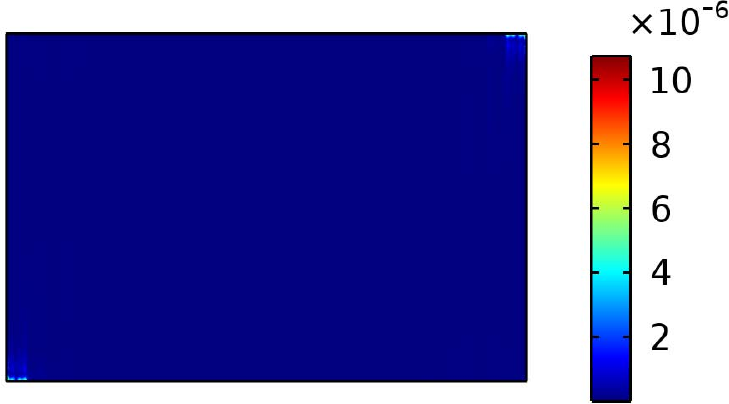}}\\
   \subfigure[]{
    \label{terecreal3}
   \includegraphics[width=0.48\columnwidth,draft=false]{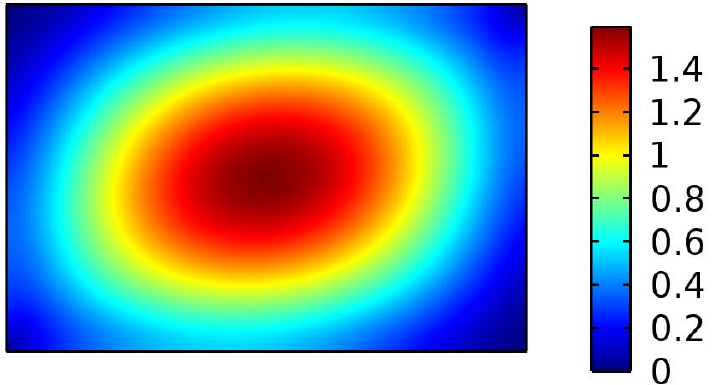}}
   \subfigure[]{
    \label{terecreal4}
   \includegraphics[width=0.48\columnwidth,draft=false]{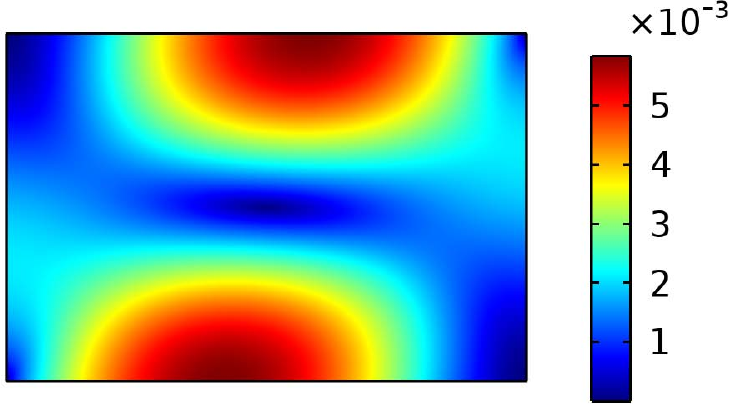}}  \\
   }
 \caption{The magnitude of the electromagnetic transverse and longitudinal components for the first mode.. $k_{z}^{(1)}=16.112$\,rad/m. The unit of the electric field amplitude is V/m.
 The unit of the magnetic field amplitude is A/m. (a) $e_{z}^{(1)}$.
          (b) $h_{z}^{(1)}$.
          (c) $\ee_{t}^{(1)}$.
          (d) $\h_{t}^{(1)}$.}
\label{mode1}
\end{figure}

\begin{figure}[ht]
  \centering
  {\subfigure[]{
    \label{terecreal1}
   \includegraphics[width=0.48\columnwidth,draft=false]{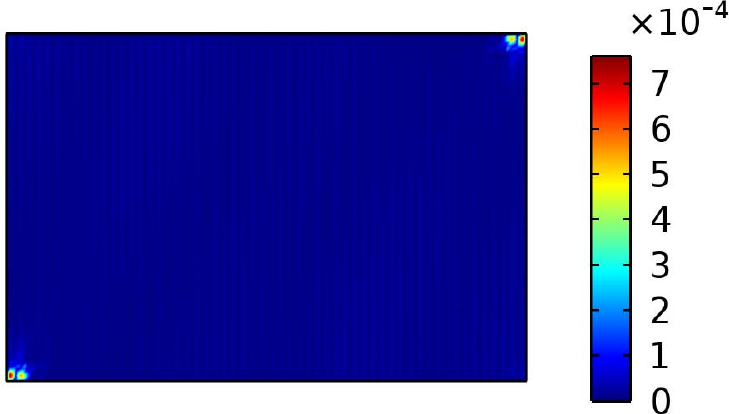}}
     \subfigure[]{
    \label{terecreal2}
   \includegraphics[width=0.48\columnwidth,draft=false]{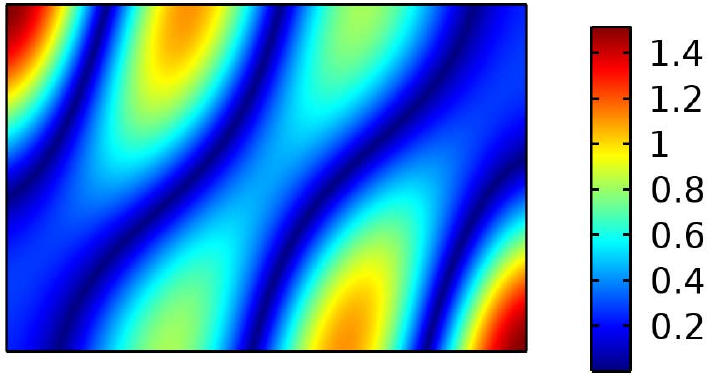}}\\
   \subfigure[]{
    \label{terecreal3}
   \includegraphics[width=0.48\columnwidth,draft=false]{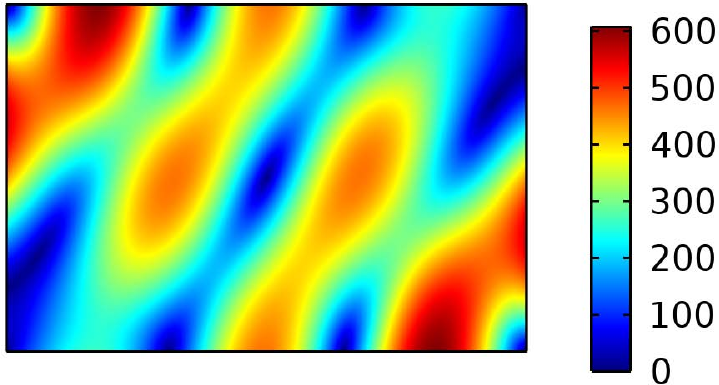}}
   \subfigure[]{
    \label{terecreal4}
   \includegraphics[width=0.48\columnwidth,draft=false]{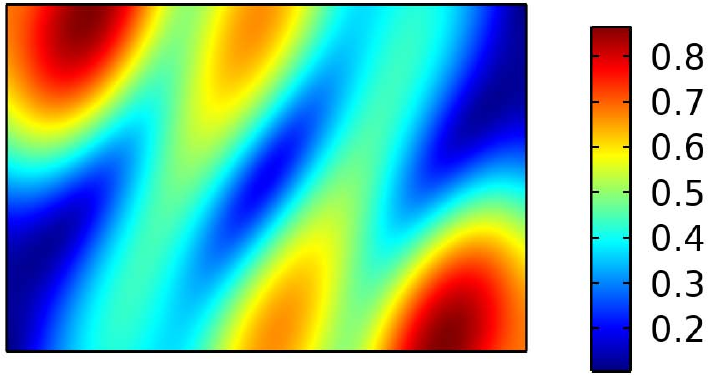}}  \\
   }
 \caption{The magnitude of the electromagnetic transverse and longitudinal components for the second mode. $k_{z}^{(2)}=28.762$\,rad/m. The unit of the electric field amplitude is V/m.
 The unit of the magnetic field amplitude is A/m. (a) $e_{z}^{(2)}$.
          (b) $h_{z}^{(2)}$.
          (c) $\ee_{t}^{(2)}$.
          (d) $\h_{t}^{(2)}$.}
\label{mode2}
\end{figure}
\noindent where $a_{ik}$, $b_{33}$, $\nu_{1}$ and $\nu_{2}$ are all nonzero real numbers,
such that $\d{\ep}$ and $\d{\mu}$ are two symmetric positive definite matrices.
It is easy to verify that (\ref{eye1})-(\ref{eye4}) holds if $\d{\ep}$ and $\d{\mu}$ are of
the form (\ref{EEE}). In next section, by a numerical experiment, we verify that there exist the pure TE modes and TM modes in the metallic waveguide filled with a homogenous, fully anisotropic and lossless medium provided that $\d{\ep}$ and $\d{\mu}$ satisfy the condition (\ref{EEE}).

%\section{Pure TE modes}
%
%
%\section{Pure TM modes}

\section{Numerical Example}
It is very difficult to directly solve PDEs (\ref{eq:1}) by a numerical method. Hence, after eliminating the magnetic field $\H$ in PDEs (\ref{eq:1}), we get the following PDEs:
\begin{subequations} \label{eqsx:1}
\begin{numcases}{}
  \curl\big(\d{\mu}^{-1}\curl\E\big)=\omega^2\d{\ep}\E\mbox{~in~}\Gamma\label{eqsx:1a}\\
  \^n\times\E={\bf{0}}\mbox{~on~}\partial\Gamma\label{eqsx:1c}
\end{numcases}
\end{subequations}
where $\E=[\ee_{t}(x,y)+\^ze_{z}(x,y)]\textrm{e}^{-\j k_{z} z}$ are unknown. Once $\E$ is obtained by a numerical method, then $\H=\j\omega^{-1}\d{\mu}^{-1}\curl\E$.
\indent Due to the length of this letter, only one numerical experiment will be considered in the section.
Assume that the cross section $\Gamma$ of the waveguide is a rectangle, whose length and width are   15\,mm $\times$ 10\,mm, respectively. Furthermore, assume that the medium parameters $\d{\ep}$ and $\d{\mu}$ of the lossless and fully anisotropic medium in the metallic waveguide are of the following form
\begin{equation}\label{AEEE}
    \d{\ep}=\begin{bmatrix}
            2&1&1\\
            1&4&1\\
           1&1&2
    \end{bmatrix}\ep_{0},\quad
        \d{\mu}=\begin{bmatrix}
            4&2&2\\
            2&8&2\\
           2&2&8
    \end{bmatrix}\mu_{0},
\end{equation}
where $\ep_{0}$ and $\mu_{0}$ are the permittivity and permeability of free space, respectively. It is clear that (\ref{AEEE}) satisfy the above condition (\ref{EEE}). Set $k_{0}=\omega\sqrt{\ep_{0}\mu_{0}}=16$\,$\textrm{m}^{-1}$.

Subject on all the numerical methods to solve PDEs (\ref{eqsx:1}), the finite element method is most suitable for solving this kind of electromagnetic eigenvalue problem. Therefore, COMSOL Multiphysics is employed to numerically simulate the problem (\ref{eqsx:1}). In COMSOL Multiphysics, we use uniform rectangular division, and divide $\Gamma$ into 532 congruent rectangles. In addition, the edge basis function of three-order on rectangular element is used to discretize $\ee_{t}$ and the nodal basis function associated with standard cubic element on rectangular element is used to discretize $e_{z}$. As seen from the numerical results of COMSOL, there are eight propagable physical modes. Here, the field distribution of the first and second propagable mode are listed in Fig. \ref{mode1} and Fig. \ref{mode2}, respectively.

From Fig.\ref{mode1}(b), it can be observed that $h_{z}^{(1)}\approx0$, which implies that the first propagable mode is just a TM mode. From Fig.\ref{mode2}(a), it can be observed that $e_{z}^{(2)}\approx0$, which implies that the second propagable mode is just a TE mode. According to the numerical results of this example, the proposed condition (\ref{EEE})  does confirm the existence of the pure TE and TM modes in the anisotropic waveguide.

%\subsection{The Simulation by $\E$-Scheme}
%
%\subsection{The Simulation by $\H$-Scheme}

%\begin{gather*}
%\d{\ep}_{r}=
%\begin{bmatrix}
%2&1&1\\
%1&4&1\\
%1&1&2
%\end{bmatrix}\quad
%\d{\mu}_{r}=
%\begin{bmatrix}
%1&0.5&0.5\\
%0.5&2&0.5\\
%0.5&0.5&2
%\end{bmatrix}
%\end{gather*}

\section{Conclusion}
If the permittivity and permeability tensors of the lossless and fully anisotropic material in the metallic waveguide is given by  (\ref{EEE}), then the pure TE and TM modes will exist in this anisotropic waveguide as long as the frequency of electromagnetic wave. This theory extends the classic electromagnetic waveguide theory.

%\bibliographystyle{IEEEtran}
%\bibliography{refnew}


%\end{IEEEbiography}

\end{document}

%% file: latex_defs.tex
\def\^{\hat}
\def\~{\tilde}
\def\3h{{3\over 2}}

\def\v#1{{\bf#1}}

\def\E{{\v E}}
\def\ee{{\v e}}
\def\h{{\v h}}

\def\H{{\v H}}

\def\F{{\v F}}

\def\N{{\v N}}

\def\ep{\epsilon}

\def\pd#1#2{{{\partial #1}\over{\partial #2}}}

\def\gradt{\nabla_{t}}

\def\curl{\nabla\times}

\def\curlt{\nabla_{t}\times}

\def\eqn#1$${\eqno{{\rm #1}}$$}

\def\ep{\epsilon}

\def\~{\tilde}

\def\^{\hat}

\def\d#1{\overline{\overline{#1}}}

\def\XXint#1#2#3{{\setbox0=\hbox{$#1{#2#3}{\int}$}
     \vcenter{\hbox{$#2#3$}}\kern-.5\wd0}}